\documentclass[aps,pra,onecolumn,10pt]{revtex4-2}
\usepackage{hyperref}
\usepackage{amsmath}
\usepackage{graphicx}
\usepackage{xcolor}
\usepackage{amssymb}
\usepackage{wrapfig}
\usepackage{geometry}
\usepackage{amssymb}
\usepackage{wrapfig}
\usepackage{geometry}
\begin{document}

\title{Vertical velocities in quasigeostrophic laboratory vortices}

\author{Marine Aulnette}
\author{Michael Le Bars}
\author{Patrice Le Gal}
\affiliation{Aix-Marseille Univ, CNRS, Centrale Med, IRPHE, Marseille, France}

\date{\today}

\begin{abstract}
\vspace{0.4 cm}
In the present study, we test the predictions of the $\omega$-Equation against laboratory experiments with direct measurements of the vertical velocity w. Our results are further completed through the use of theoretical models of oceanic vortices, with the aim of helping oceanographers in better quantifying regions of upwelling and downwelling in the ocean. Using a rotating table and density stratification, we investigate non-axisymmetric surface vortices. The predicted vertical velocities calculated from the $\omega$-Equation are relatively small (w $\simeq \pm 20 \mu$m·s$^{-1}$) and primarily appear at the vortex edges, where the vorticity sign changes, acting to restore flow stratification. However, our estimates of w, obtained from the divergence of the horizontal velocity field measured by PIV, are five times larger. This discrepancy is further confirmed by direct particle tracking measurements, which indicate a magnitude of approximately $\pm 100$ $\mu$m·s$^{-1}$ for w. To address this inconsistency, we incorporate dissipative terms into the $\omega$-Equation to assess the role of viscous diffusion in enhancing internal recirculation in the vortex and thus vertical velocity magnitude. This hypothesis is favorably tested on a Gaussian vortex model.
\end{abstract}

\maketitle

\section{Introduction}

Large oceanic structures subjected to Earth's rotation are driven by geostrophic equilibrium which allows large scale oceanographic models to successfully reproduce the near two dimensional dynamics of large eddies and currents. However, smaller-scale structures such as fronts and filaments differ from this description. They generate three dimensional weak movements that do not respect the constraint of geostrophic equilibrium. These appear as part of the ageostrophic response to restore the geostrophic balance of unstable structures. These 3D motions thus possess vertical velocity fields, which even if they have small amplitude, are tremendously important for balancing the global fluxes of heat, scalars and biochemical components in the oceans. Knowing and parametrizing correctly, on the basis of realistic physical models, these small ageostrophic movements is one of the major oceanographers' challenges for understanding the oceans fine-scale dynamics. 

Mesoscale and submesoscale dynamics correspond to three-dimensional oceanic processes, spanning horizontal length scales from 1 to 100 km and evolving over timescales from hours to days \cite{mcwilliams2016submesoscale}. Mesoscale geophysical flows are dominantly horizontal. Horizontal velocities are several orders of magnitude larger than vertical ones. Even so, mesoscale vertical motions are also important for the marine ecosystem, since they transport large volumes of oceanic waters. But, direct experimental measurements of these vertical velocities are difficult to carry in-situ. They are of the order of $1$ mm.s$^{-1}$ or less, placing them below the noise level of many measuring devices currently available. In the context of atmospheric applications, indirect methods, such as the $\omega$-Equation, have been developed to infer these vertical velocities from horizontal flow and more readily measurable quantities such as density \cite{panofsky1946methods,sherman1953estimates}. After several versions of the equation, Hoskins et al. proposed a new refined quasigeostrophic (QG) ``Q-vector" $\omega-$Equation, used in this study \cite{hoskins1978new}. The model states that small vertical motions w may arise to re-establish the thermal wind balance. This quasigeostrophic version of this equation has been used for decades and is, by construction, better suited to the study of low Rossby numbers (that compares inertia to rotation) configurations, away from intense vertical velocities. This classical QG $\omega-$Equation includes a source term called the ``frontogenetical vector" that describes the geostrophic advection of the density gradients. Frontogenesis describes the ageostrophic circulation that arises when density fronts strengthen \cite{lapeyre2006oceanic,mcwilliams2021oceanic}. Generalization of this $\omega-$Equation, beyond the quasigeostrophic assumption, have since been developed \cite{viudez1996nature,giordani2006advanced,pietri500skills}. Numerous studies of oceanic vertical velocities have been carried out using the $\omega-$Equation. Most of them have been used on in-situ observations where 2D horizontal measurements serve as input to predict the vertical velocity component  \cite{rousselet2019vertical,tzortzis2021impact}. Some have confronted results from this model to independent measurements of the vertical motions to assess the accuracy of the $\omega-$Equation and have concluded that the $\omega-$Equation has reasonable skills at the mesoscale (see references in \cite{pietri500skills}). Finally, some groups have applied the $\omega-$Equation to oceanic numerical models outputs \cite{mason2017subregional,uchida2019contribution} or even in numerical idealized settings \cite{viudez2003vertical,viudez2004potential,viudez2018two}. In general, theses studies show that the basic QG form of the $\omega-$Equation is not sufficient to fully predict the vertical velocities. A more exhaustive overview of the literature is proposed in \cite{pietri500skills}. As shown in this last study, estimations of the vertical velocity obtained from a generalized form of the $\omega-$Equation are systematically lower by a factor 2 to 100 than their oceanographic model outputs.
To the best of our knowledge, no study has ever tested the $\omega-$Equation predictions against laboratory experiments with direct measurements of the vertical velocity.

In the present study, we propose a comparison between vertical velocities experimentally measured and those inferred with the $\omega-$Equation in laboratory vortices. We focus here on vortical structures, as they are ubiquitous in nature and contribute to climate equilibrium on Earth by transporting heat or momentum inside the oceans or at their surface. Geophysical oceanic vortex dynamics is mainly determined by the Earth's rotation and ocean density stratification, and their horizontal scales are usually much larger than their vertical ones \cite{van2009laboratory}. For vortices in a stratified flow or for floating vortices, numerous laboratory experiments have been performed to characterize their aspect ratios \cite{aubert2012universal,de2017laboratory}, time evolution ~\cite{facchini2016lifetime}, instabilities~\cite{griffiths1981stability} and interactions ~\cite{griffiths1987coalescing,orozco2020coalescence}. It was found that their aspect ratio is a function of the Rossy number and the Coriolis parameter, that characterize respectively the rotation of the flow and the Earth, and the Brünt-Väisälä frequency, that characterizes the density stratification of the system, both in and out of the vortex. In the case of a strong stratification, the vortex is more constrained vertically and spreads horizontally. In the case of a strong rotation, the vortex is elongated vertically. By studying the time evolution of such structures, Facchini and Le Bars \cite{facchini2016lifetime} highlighted in their model that viscous diffusion triggers secondary vertical recirculations which play a crucial role in the time evolution of vortices.

For the purpose of this study, we generate several anticyclonic vortices at once by the injection of a finite volume of pure water at the surface of a salted water rotating bulk, similarly to \cite{griffiths1981stability,aubert2012universal,de2017laboratory}. Careful measurements of the density, with laser induced fluorescence, and of velocity fields, with particle image velocimetry, are performed in order to recover the necessary data to apply the $\omega-$Equation to our system. Two methods are used to reconstruct the vertical velocity field using the divergence of the horizontal velocity and tracer particle tracking. The experimental vertical velocity is larger than the $\omega-$Equation prediction. A new version of the $\omega-$Equation, including viscous and salt diffusion, is derived and opens new perspectives on this problem. 

The paper is structured as follows. Section~\ref{sec:omega} is dedicated to establishing the $\omega-$Equation model as well as testing it on two vortex models (Gaussian \cite{viudez2018two} and shielded) and showing some examples of use on real oceanic vortical structure. In section~\ref{sec:experiments}, we propose a comparison between vertical velocities w inferred from the $\omega-$Equation and direct measurements in our laboratory vortex. The paper ends with a summary of our main findings in section~\ref{sec:conclusion} and presents new perspectives.

\section{Introducing the \texorpdfstring{$\omega - $}Equation and testing vortex models} \label{sec:omega}
In the following, we consider a quasi-geostrophic flow such as the Rossby number Ro, that compares inertia to rotation, is small:  $\textup{Ro} = U / Lf \ll 1$ with $U$ a characteristic velocity, $L$ a characteristic length and $f$ the Coriolis parameter. We can perform an expansion of the velocity field in terms of Ro. The velocity field is as follows :

\begin{equation}
    \vec{u} = \vec{u_g} + \textup{Ro} \; \vec{u_a},
    \label{eq:expansion}
\end{equation}

with $\vec{u_g}=(u_g,v_g,0)$ the geostrophic horizontal component of the velocity and $\vec{u_a} = (u_a,v_a,\textup{w})$ its first order correction in Ro. As classically done, the expansion for the pressure field $p$ is not performed as it is considered in geostrophic balance in the horizontal directions and in hydrostatic balance in the vertical direction. The Navier-Stokes equations in the Boussinesq approximation read :
\begin{equation}
    \frac{\partial}{\partial t} \vec{u} + (\vec{u} \cdot \vec{\nabla}) \vec{u} = -\frac{1}{\rho_0} \frac{\partial p}{\partial z} \vec{z} - \frac{\rho}{\rho_0 } g \vec{z} - f \vec{z} \times \vec{u},
    \label{eq:NS_momentum}
\end{equation} 
\begin{equation}
    \frac{\partial \rho}{\partial t} + (\vec{u} \cdot \vec{\nabla}) \rho + \rho_0 \vec{\nabla} \cdot \vec{u} = 0,
    \label{eq:NS_continuity}
\end{equation}
\begin{equation}
    \nabla \cdot \vec{u}  = 0,
    \label{eq:divergence}
\end{equation}
with $\rho$ the density ($\rho_0$ the reference density) and $g$ the acceleration of gravity. Viscosity is neglected for the moment as Reynolds numbers are usually very high in the ocean.  
At zero order in Rossby number, equation~\ref{eq:NS_momentum} gives the geostrophic and hydrostatic balances. At first order in Rossby number, the set of equations~\ref{eq:NS_momentum}$-$\ref{eq:divergence}, with the Brunt-Väisälä frequency $N^2 = -\frac{g}{\rho_0}  \frac{\partial \rho}{\partial z}$ and $\vec{\nabla}_h$ the horizontal gradients, reduces to:
\begin{equation}
    \frac{\partial}{\partial t} \vec{u_g} + (\vec{u_g} \cdot \vec{\nabla}_h) \vec{u_g} = -\frac{1}{\rho_0} \frac{\partial p}{\partial z} \vec{z} - \frac{\rho}{\rho_0 } g \vec{z} - f \vec{z} \times \vec{u_a},
    \label{eq:NS_momentum_order1Ro}
\end{equation}

\begin{equation}
    \frac{\partial \rho}{\partial t} + (\vec{u}_g \cdot \vec{\nabla}_h) \rho - \frac{\rho_0}{g} N^2\textup{w} = 0,
    \label{eq:NS_continuity_order1Ro}
\end{equation}

Together, these equations at first order in Rossby number reduce to the so called $\omega - $Equation \cite{hoskins1978new,viudez2018two} :

\begin{equation}
 N^2 \nabla^2_h \textup{w} + f^2 \frac{\partial^2}{\partial z^2} \textup{w} = 2 \frac{g}{\rho_0} \vec{\nabla}_h \cdot (\vec{\nabla}_h \vec{u_g} \cdot \vec{\nabla}_h \rho). 
  \label{eq:omega_equation}
\end{equation}

Equation~\ref{eq:omega_equation} is a generalized Poisson equation with a modified Laplacian operator applied to the vertical velocity magnitude w. The source term depends on the horizontal gradient of the density and on the geostrophic velocity field. This equation states that vertical velocities will arise when the gradients of density are advected by the geostrophic flow. Note that the source term on the right hand side of equation~\ref{eq:omega_equation} vanishes if the geostrophic flow $\vec{u}_g$ and the density gradients are orthogonal. The $\omega-$Equation is presented here in its simplest form. Additional source terms may be considered given the configuration under study. Since the present version of the $\omega-$Equation is better suited by nature to low Rossby number regions, i.e. away from intense vertical velocity regions, a generalized formulation of equation~\ref{eq:omega_equation} was derived to hold even in high Rossby number areas \cite{viudez1996nature,viudez2004potential}. This expanded formulation highlights additional vertical velocity forcing processes such as kinetic deformation that arises in shear and confluence situations, and mixing and momentum diffusion which can also disrupt the thermal wind balance \cite{pietri500skills}.

To solve equation~\ref{eq:omega_equation}, we write the stretched quasigeostrophic vertical coordinate $\tilde{z} = \frac{N}{f} z$. The left hand term becomes a simple Laplacian operator in the quasigeostrophic space. Finally, equation~\ref{eq:omega_equation} reads $\tilde{\nabla}^2 \textup{w} = S$ with $S$ the source term depending on the horizontal variation of the geostrophic velocity and density fields. Then, we use a finite difference algorithm to inverse the Laplacian operator. 

\subsection{Predictions from oceanographic data}
Oceanographers use the $\omega-$Equation on in-situ data or on reconstructed synthetic fields obtained from satellite measurements and models. For example, Nardelli \cite{nardelli2013vortex} used the $\omega-$Equation model to diagnose vertical velocities in a mesoscale cyclonic eddy located near South Africa in the Agulhas current. A pattern of alternating upward and downward motions on the edges of the vortex is calculated. These azimuthal oscillations of the vertical velocity are said to be compatible with potential vorticity anomalies propagating along the radial gradient of potential vorticity, also known as Vortex Rossby Waves. Another example of the use of the $\omega-$Equation on data obtained from satellite measurements includes the collection of mesoscale eddies located at the Brazil-Malvinas Confluence \cite{mason2017subregional}. In most of the studied eddies, the predominant pattern of quasigeostrophic motion is a dipole of upward and downward vertical velocity, consistent with the theory of conservation of potential vorticity. Finally, a more refined version of the $\omega-$Equation is used on in-situ measurements performed near the Canary Islands by Barcel\'o-Llull et al. \cite{barcelo2017ageostrophic}. Once again, a similar pattern of upwelling and downwelling is diagnosed at the eddy periphery. The common feature in these examples is the slight ellipticity of each mesoscale eddy, emphasizing that the vortex lack of axisymmetry is a key element to obtain vertical velocity of typical order of magnitude $10^{-4}$ to $10^{-3}$ smaller than the horizontal velocity. Indeed, in a non-axisymmetric structure, the geostrophic flow is not perpendicular to the density gradients, as required to have a non-zero source term in equation~\ref{eq:omega_equation}. 

\subsection{Testing vortex models}
In this section, we propose to infer vertical velocities w using equation~\ref{eq:omega_equation} on two vortex models, similarly to Viudez's analysis of vertical velocity modes in a Gaussian vortex \cite{viudez2018two}. His aim was to find the most simple vertical velocity patterns associated with mesoscale eddies. For both models, the geostrophic velocity field of the vortex is based on a geopotential $\phi$, that will depend on the model considered, and will be given in a cylindrical frame of reference $(r,\theta,z)$ by: 

$$\vec{u}_g = v_m \frac{r}{r_m} \phi(r) \quad \vec{e}_\theta, $$
with $v_m$ the maximal magnitude of velocity located at radius $r_m$. The associated density profile is deduced using the combination of the geostrophic and hydrostatic balances, i.e. the thermal wind balance
\begin{equation}
 \vec{\nabla}_h \rho = \frac{\rho_0 f}{g} \vec{z} \times \frac{\partial \vec{u}_g}{\partial z}. 
 \label{eq:thermal_wind_balance}
 \end{equation}

In the following we use as a control parameter the Rossby number defined as $\textup{Ro} = \omega_c / 2f$ with $\omega_c = v_m / r_m$ the vortex core vorticity and $f$ the Coriolis parameter.

\subsubsection{Gaussian vortex model}
\begin{figure}[t]
\centering
\includegraphics[width=13cm]{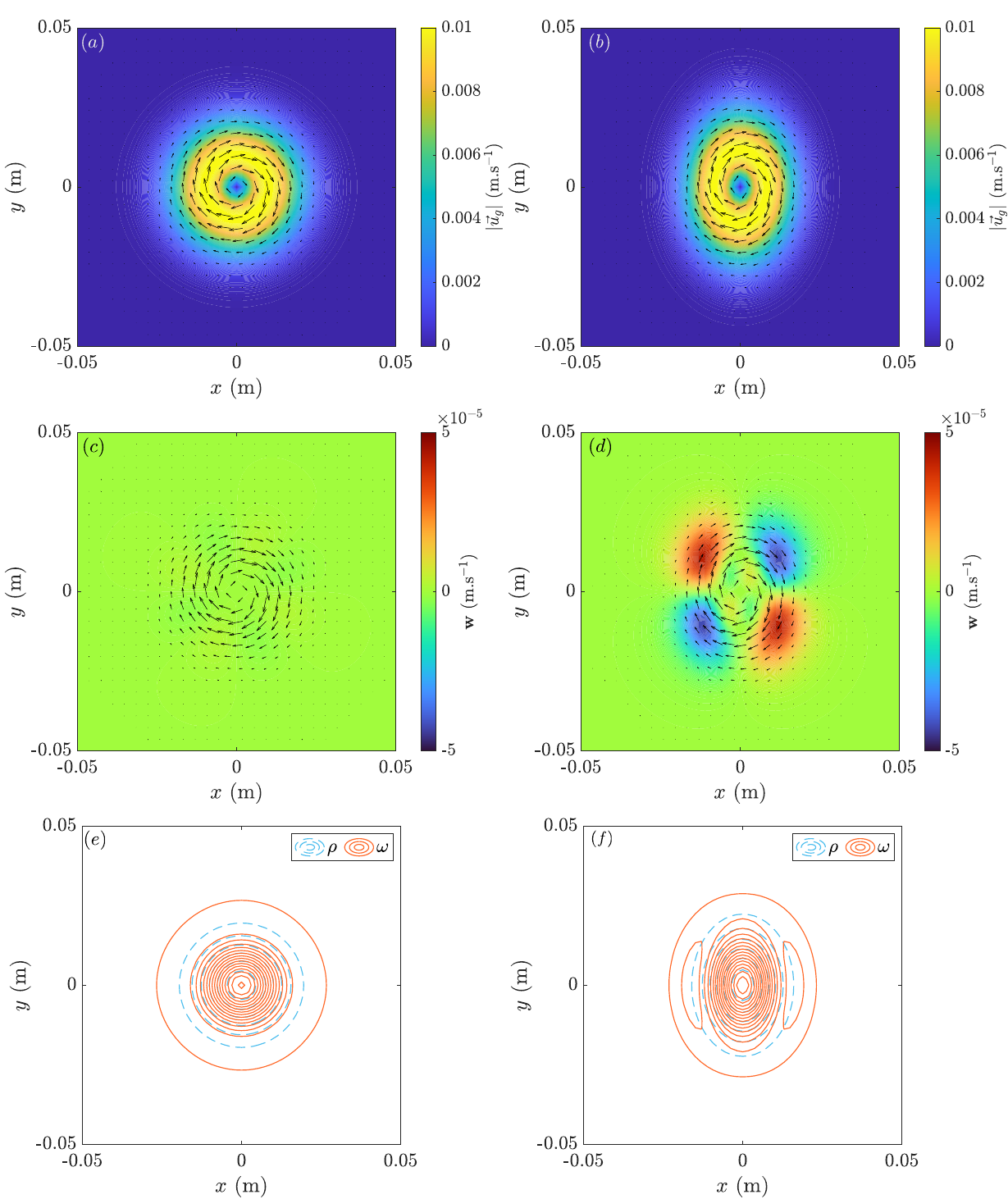}
\caption{For $\textup{Ro} = -0.3$, $r_m = 2 \times 10^{-2}$ m.s$^{-1}$, $f = 2$ s$^{-1}$ and $\rho_0 = 1005$ kg.m$^{-3}$: diagnosis of the vertical velocity w of a Gaussian vortex model with the $\omega-$equation at $\tilde{z} \simeq -0.002$ (eq. \ref{eq:omega_equation}). (a) Horizontal velocity field $\vec{u_g}$ for $\beta = 1$ (axisymmetric configuration). (b) Horizontal velocity field $\vec{u_g}$ for $\beta = 1.4$ (similar horizontal aspect ratio as Viudez's study \cite{viudez2018two}). (c,d) Vertical velocity w resulting from the $\omega-$equation for the axisymmetric vortex and the elliptical vortex. (e,f) Isocontours of density and vorticity for both axisymmetric and elliptical vortices. In panel (e), both fields are parallel and no vertical velocity is generated as can be seen in panel (c). On the contrary, in panel (f), both field contours intersect and vertical velocities are generated as shown in panel (d).}
\label{fig:figure1}
\end{figure}

Following Viudez \cite{viudez2018two}, we consider a classical Gaussian vortex based on the geopotential $\phi$

$$
   \phi(x,y,\tilde{z}) = \exp \left\{ - \left( \frac{x^2}{A_x^2} + \frac{y^2}{A_y^2} + \frac{\tilde{z}^2}{A_{\tilde{z}}^2}\right)\right\}
$$
with $(A_x,A_y,A_{\tilde{z}})$ the semi-axes of the vortex. We define the horizontal aspect ratio $\beta = A_y / A_x$. The horizontal semi-axes can be derived from $r_m$ with $r_m = (A_x^2 + A_y^2)^{1/2}$. Therefore, we obtain:

$$A_x = \frac{r_m}{\sqrt{1 + \beta^2}} \text{ and } A_y = \frac{\beta r_m}{\sqrt{1 + \beta^2}}.$$  

The vertical aspect ratio of vortices in rotating stratified flows has been well studied in the literature, for laboratory and numerical vortices as well as for oceanic structures. It is given by $\alpha = A_z / 2 r_m = -f^2 \textup{Ro} / N^2$ with $N^2 = - 4 g \Delta \rho / \rho_0 r_m$ the Brünt-Väisälä frequency \cite{hassanzadeh2012universal,aubert2012universal,de2017laboratory}. We derive our vertical semi-axis $A_z$ from $\alpha$  

$$A_z = 2 r_m \left(-\frac{f^2 \textup{Ro}}{N^2}\right).$$

The associated geostrophic velocity $\vec{u}_g$ field (figure~\ref{fig:figure1}, a \& b) and vorticity $\omega$ field (iscontours of $\omega$ are shown in figure~\ref{fig:figure1}, e \& f) are respectively given by

$$\vec{u}_g = v_m  \frac{y}{A_y}  \phi \; \vec{e}_x - v_m  \frac{x}{A_x}  \phi \; \vec{e}_y,$$

and $$\omega(x,y,\tilde{z}) = v_m \left[ \frac{2x^2-A_x}{A_x^2} + \frac{2y^2-A_y}{A_y^2} \right] \phi,$$
with $v_m = \omega_c r_m = - 2 r_m f \textup{Ro}$ the maximum velocity magnitude. 
The corresponding density profile is calculated from the thermal wind balance equation~\ref{eq:thermal_wind_balance}:

$$ \rho(x,y,\tilde{z}) = C(x,y,z) + \frac{\rho_0 f}{g} v_m (A_x + A_y) \frac{\tilde{z}}{A_{\tilde{z}}^2}  \phi, $$
with $C(x,y,z)$ the background density field that could be stratified or homogeneous. In our case, we chose an homogeneous background density $C(x,y,z) = \rho_0 = 1005$ kg.m$^{-3}$, corresponding to the density of salted water in our experiments presented in the next section.

\bigbreak
In Viudez's study, it is highlighted that vertical velocities arise in two specific configurations. For an axisymmetric vortex ($A_x = A_y = A_{\tilde{z}}$), geostrophic balance is not broken : the source term of the $\omega-$Equation is zero because all horizontal variations vanish due to symmetry. Therefore, no vertical velocity appear for axisymmetric vortices (figure~\ref{fig:figure1} (c)). However, if the shape is slightly altered, as it is the case for elliptical vortices for instance $(A_x \neq A_y)$, a multipolar pattern of upward and downward motions appears along the edges of the vortex (see figure~\ref{fig:figure1} (d)). This pattern occurs in order to conserve the potential vorticity anomaly along the non-axisymmetric trajectories of fluid particles. Potential vorticity takes into account both vorticity and distribution of matter around the rotation axis. Therefore, if the distribution is not axisymmetric, and the rotation rate stays the same, the column of matter stretches along the rotation axis, i.e. vertical motion is triggered. Figure~\ref{fig:figure1} panels (e) and (f) also illustrate that, contrary to the case of an axisymmetric vortex, isodensity and vorticity contours cross in the case of a non axisymmetric vortex, producing a source term in the $\omega-$Equation in each quadrant. Note finally that vertical motion is also forced if the rotation axis of the circular vortex is tilted relative to the gravity direction, but this configuration, studied in depth by Viudez, will not be further developped here \cite{viudez2018two}. 

Figure~\ref{fig:figure1} also shows that for horizontal velocities of the order of 1 cm.s$^{-1}$, vertical velocities are expected to be close to 10 $\mu$m.s$^{-1}$. The ratio w$/u_g$ is $1/1000$, confirming that ageostrophic vertical velocity are particularly small compared to the horizontal currents.

\subsubsection{Shielded vortex model} \label{sec:shielded}
\begin{figure}[t]
\centering
\includegraphics[width=14cm]{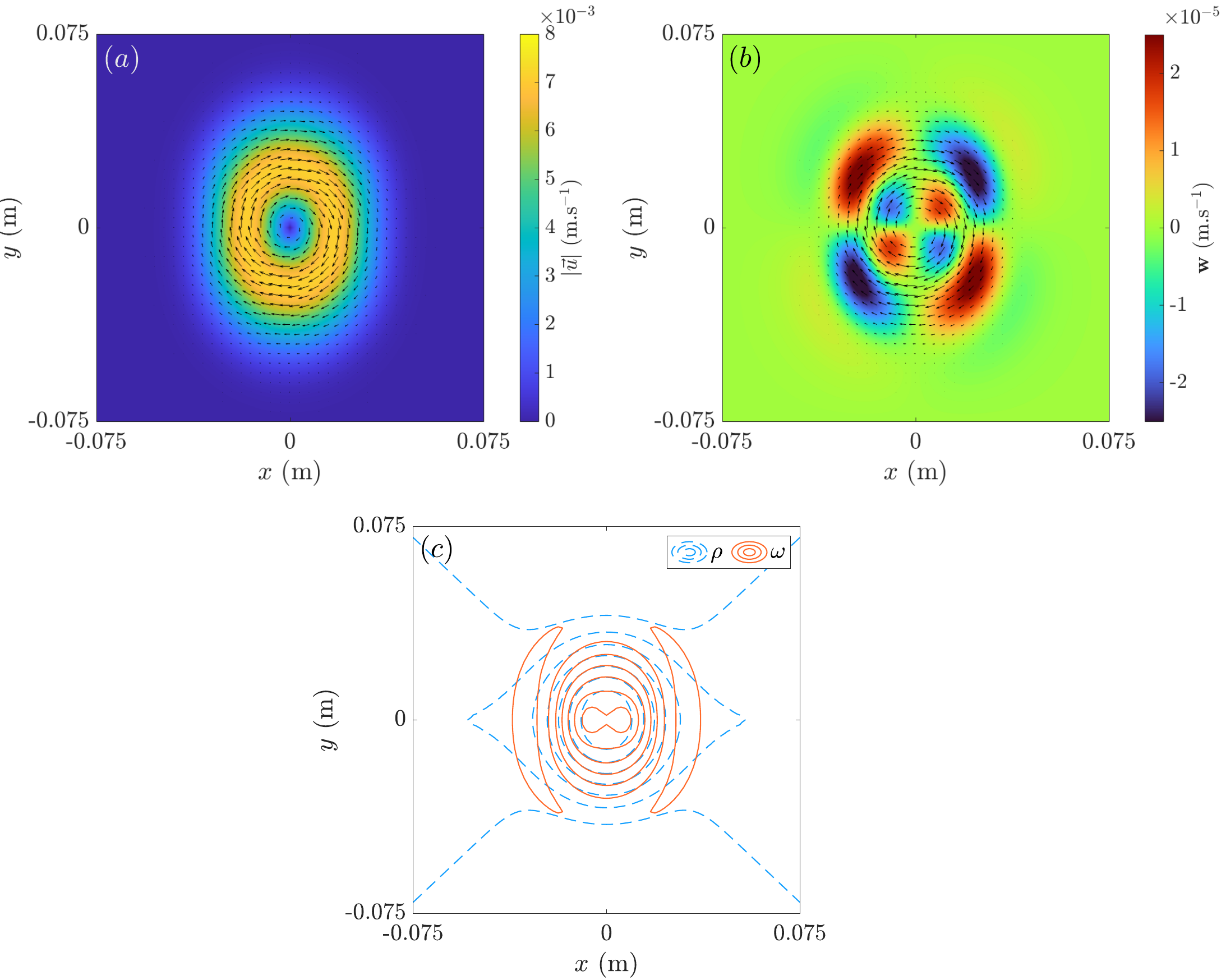}
\caption{For $\textup{Ro} = -0.3$, $r_m = 2 \times 10^{-2}$ m, $f = 2$ s$^{-1}$ and $b = 3$. (a) Shielded vortex velocity field below the free-surface. (b) Vertical velocity w diagnosed with the $\omega-$Equation from the shielded vortex model. (c) Isocontours of density and vorticity.: their intersections correspond to where vertical velocities arise.}
\label{fig:figure2}
\end{figure}

Although a Gaussian model is reasonable to characterize vortices, oceanic eddies exhibit a slightly more complex shape. Indeed for some structures, vorticity is maximum at the core but a ring of opposite sign vorticity surrounds this maximum, forming a shield between the inside and the outside of the vortex \cite{kloosterziel1992evolution,carton1994life}. In the literature, the shielded vortex model has been used to describe atmospheric vortices of Earth and Jupiter \cite{chan1987analytical, demaria1984comments, li2020modeling} as well as oceanic vortices such as those in the North Brasil Current \cite{castelao2011sea}. 

For the purpose of our analysis, we will compare and contrast the $\omega-$Equation results obtained for the classical Gaussian model and for the shielded vortex model. Since we want to study the ageostrophic circulation that arises in these vortices, we assume that the shielded vortex velocity profile decays following a Gaussian trend in the vertical direction. Thus, we define the shielded geopotential $\phi$:

$$\phi(r,\tilde{z}) = \exp \left[ \frac{1}{b} \left(1 - \left(\frac{r}{r_m} \right)^b \right) - \frac{\tilde{z}^2}{A_{\tilde{z}}^2} \right].$$

The velocity field and the vorticity, respectively, are given by

\begin{equation}
    u_{\theta}(r,\tilde{z}) =  v_m  \left( \frac{r}{r_m} \right) \phi,
    \label{eq:shielded_velocity2}
\end{equation}
and
\begin{equation}
    \omega(r,\tilde{z}) = \frac{v_m}{r_m} \left( 1 - \frac{1}{2} \left( \frac{r}{r_m}  \right)^b \right) \phi,
    \label{eq:shielded_vorticity2}
\end{equation}

with $v_m = \omega_c r_m = 2 r_m f \textup{Ro}  \exp(1/b)$ the maximum tangential velocity, $r_m(\theta)$ the radial location of the tangential velocity maximum, and $b$ determines the rate at which the tangential velocity decays with $r$. The intensity of the shield can be tuned by changing this parameter $b$. As in the previous test case, $A_z$ is the vertical dimension of the vortex and is derived from the aspect ratio $\alpha$. 

The source term of the $\omega-$Equation requires a contribution from the horizontal gradients of density. Because we consider a quasigeostrophic flow, we assume that the matching density profile to our shielded vortex is, again, derived from the thermal wind balance (equation~\ref{eq:thermal_wind_balance}). The integration of this equation gives the following density field:

$$ \rho(r,\tilde{z}) = C(r,\tilde{z}) - v_m \frac{\rho_0 f}{g} \frac{2\tilde{z}}{A_{\tilde{z}}^2} \exp\left(-\frac{\tilde{z}^2}{A_{\tilde{z}}^2}\right) \exp\left(\frac{1}{b}\right) \frac{r^2}{r_mb} \left[ \frac{1}{b} \left(\frac{r}{r_m}\right)^{-2/b}  \right] \Gamma\left(\frac{2}{b} , \frac{1}{b} \left(\frac{r}{r_m}\right)^b\right),$$
with $\Gamma(a,x)$ the incomplete gamma function and $C(r,\tilde{z})$ the background density that can be stratified or constant.

 \bigbreak
 
\begin{figure}[t]
\centering
\includegraphics[width=14cm]{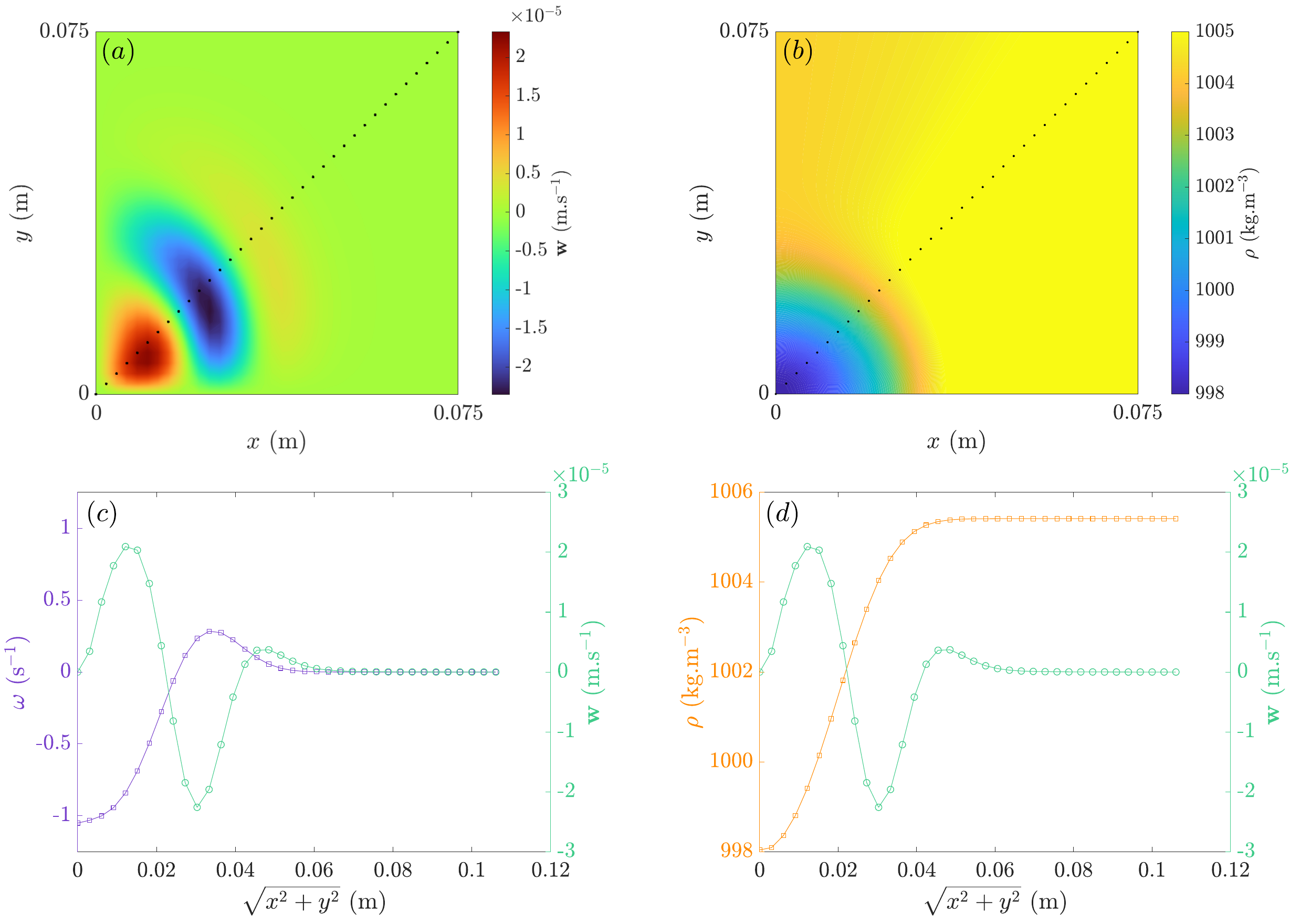}
\caption{For $\textup{Ro} = -0.3$, $r_m = 2 \times 10^{-2}$ m.s$^{-1}$, $f = 2$ s$^{-1}$ and $b = 3$. (a) Vertical velocity w diagnosed with the $\omega-$Equation from the shielded vortex model for $(x,y)>0$. (b) Fluid density derived from the thermal wind balance and the velocity field (eq.~\ref{eq:shielded_velocity2}). (c) Profile of vertical velocity (right axis) superimposed with the vorticity profile (left axis) along $x = y$ (shown as a dotted line in (a)). (d) Profile of vertical velocity (right axis) superimposed with the density profile (left axis) along $x = y$ (shown as a dotted line in (b)).}
\label{fig:figure3}
\end{figure}

Results from the $\omega-$Equation are shown in figures~\ref{fig:figure2} and~\ref{fig:figure3}. Similarly to the Gaussian model, no velocity arises if the vortex is axisymmetric. However, by making it elliptical, vertical velocities are triggered where contours of density and vorticity cross (figure~\ref{fig:figure2} (c)). In the case of shielded vortices, the expected spatial distribution of w differs from the Gaussian vortex presented in figure~\ref{fig:figure1}. We recover multipoles of w with a similar spatial location to that of the Gaussian vortex mode. However, each lobe is split into two oppositely signed parts. The direction of vertical velocity changes when the vorticity switch from negative to positive as illustrated with the profiles of w and $\omega$ plotted in figure~\ref{fig:figure3} (c). Vertical velocity location also coincides with the front between the light and dense fluid masses (figure~\ref{fig:figure3} (d)). 
 Therefore, the presence of an ageostrophic recirculation can be interpreted as a consequence of frontogenesis as classically evoked in oceanography \cite{mahadevan2016impact}.

\bigbreak 

\bigbreak
In conclusion, the $\omega-$Equation is a theoretical model used to infer vertical velocity in quasigeostrophic flows from horizontal measurable data. This equation shows that vertical motion may arise in non-axisymmetric configurations where density gradients need to be balanced. In this section, we have shown examples of oceanic eddies in the literature and have focused on vertical velocities in vortices by examining two classical vortex models. In the next section, the aim is to perform laboratory experiments creating anticyclonic floating vortices (lenses) and compare direct measurements of w to the $\omega-$Equation estimation.

\begin{figure}[t]
\centering
\includegraphics[width=13cm]{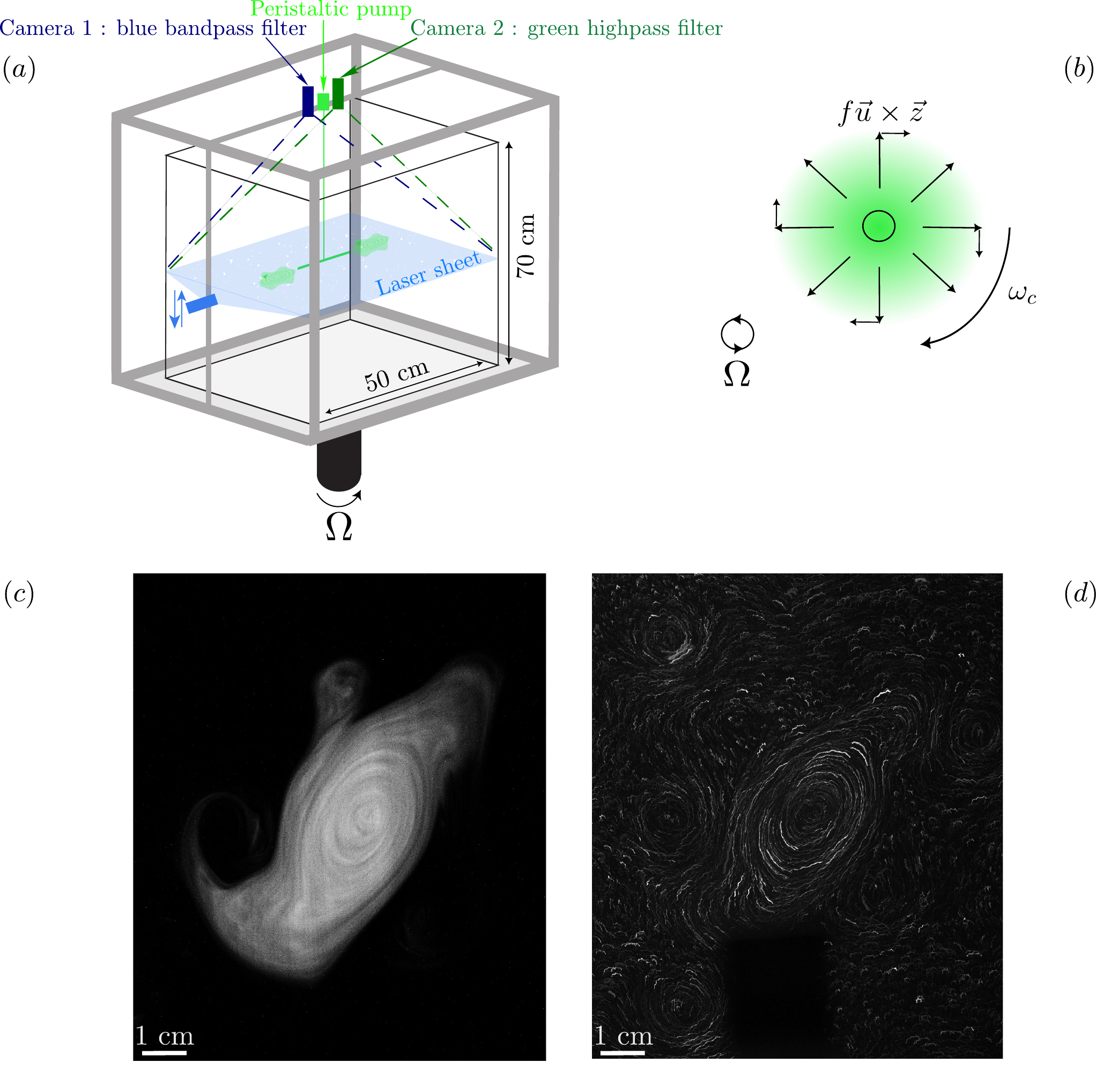}
\caption{(a) Sketch of the experimental set-up. A tank, filled with salted water seeded with silver coated PIV particles, is placed on a rotating table of rotation frequency $\Omega$. Pure water dyed with fluoresceine is injected with a peristaltic pump at the center of the free surface. A translating 488 nm laser sheet illuminates an horizontal plane of interest at fixed $z$. Two cameras mounted with bandpass blue filter and highpass green filter respectively used for PIV and LIF capture the experiment from above. (b) Process of vortex generation : the radial movement of the injection paired with the Coriolis force generates an anticyclonic vortex. (c) Raw image of Laser Induced Fluorescence : in bright, we visualize the dyed fresh water of the anticyclone surrounded by salted water in dark. (d) Superimposition of 50 raw images of PIV particles.}
\label{fig:figure4}
\end{figure}

\section{Comparing the \texorpdfstring{$\omega - $}Equation to experiments} \label{sec:experiments}
In this section, we present a comparison between experimental measurements of w and the $\omega-$Equation estimation to assess the reliability of this model for Rossby number between -0.3 and -0.1. Note that Viudez and Dritschel conducted a similar analysis but from numerical simulations of the vertical velocity field in a single mesoscale baroclinic oceanic gyre \cite{viudez2003vertical, viudez2004potential}. The time evolution of this vortex was studied by integrating the horizontal ageostrophic vorticity and conserving the potential vorticity via contour advection on isopycnal surfaces. The authors consider the generalized $\omega-$Equation but show that, in their study cases, it reduces to equation~\ref{eq:omega_equation}. Their results show that the magnitude of the $\omega-$Equation prediction is usually smaller than their simulations and depends on the vorticity magnitude of the vortex. For a vortex with a Rossby number of $-0.5$, the prediction differs of about 22$\%$ from the simulation. This gap reduces to $10\%$ for $\textup{Ro} = -0.3$. As we will see later, the difference between the $\omega-$Equation predictions and our experimental observations is much larger, for similar Rossby numbers.

\subsection{Methods and observations} \label{sec:setup}

Experiments are performed in a rectangular $50 \times 50 \times 70$ cm$^3$ tank placed on a rotating table of angular velocity $\Omega$ (illustrated in figure~\ref{fig:figure4} (a), same experimental set-up as \cite{aubert2012universal}). The Coriolis parameter of the experiment $f = 2 \Omega$ is adjusted in the range $0.5 - 2$ s$^{-1}$ with an accuracy of $0.5\%$. The tank is filled with salted water of density $\rho_f$, measured with an Anton-Paar densimeter for all the experiments performed and is equal to $\rho_f = 1005$ kg.m$^{-3}$. 

Before each experimental run, we wait for several hours after starting the rotation in order to reach full solid body rotation in the bulk. The floating vortices, or lenses, are generated by the injection of a small volume of pure water $\rho_v = 998$ kg.m$^{-3}$ at the free surface. The injection is initiated with a peristaltic pump and lasts about 5 seconds (corresponding to a volume of about $5$ ml of fresh water). During this lapse of time, fluid is injected through a T-junction orthogonal to the rotation axis. The injection system is stationary in the rotating frame.

Figure~\ref{fig:figure4} (b) illustrates the process of generating a floating vortex: the radial movement of the injection paired with the Coriolis force due to the solid body rotation initiates a rotating movement of the fresh water in the opposite direction, thus creating an anticyclonic vortex \cite{griffiths1981stability,facchini2016lifetime,de2017laboratory}. In order to avoid creating circular vortices and to actually trigger vertical motions, we chose to generate simultaneously several anticyclones \cite{griffiths1987coalescing, orozco2020coalescence}. The distance between structures is carefully chosen: they are not too close so they do not coalesce and not too far so they do not migrate away from each other. At the right separation (for two vortices, the distance between their core is 10 cm, that is about twice their typical size), their interactions only alter their shape, so they become slightly elliptical. For instance, one of these vortices is presented in figure~\ref{fig:figure4} (c) and (d).

Both salted water and fresh water are seeded with silver-coated 10 microns hollow glass particles so that the horizontal velocity field can be measured with PIV. For this, a blue laser sheet (488 nm) is embarked on the rotating table and motorized to translate in the $z$ direction to scan the vortex in height. A first camera, also embarked on the table and fixed on top of the tank (see figure~\ref{fig:figure4}), is equipped with a bandpass filter centered around 488 nm in order to isolate the tracer particles (figure~\ref{fig:figure4} (d)). The Matlab toolbox PIVlab is used to complete the velocity field calculations \cite{thielicke2021particle}. The fresh water is dyed with fluoresceine at a small concentration $C_0 = 6.4$ $\mu$g/L. Fluoresceine is a fluorescent compound with an excitation peak around 490 nm and an emission peak around 520 nm. Our blue laser sheet makes the pure water fluorescent in contrast with the salted water that remains transparent. A second camera equipped with a highpass filter, with the cut-off wavelength at 510 nm, is fixed on top of the tank to image the fluorescence (figure~\ref{fig:figure4} (c)). Laser Induced Fluorescence (LIF)  is used to measure density of the fluid: local fluorescein concentration $C(x,y,t)$ is deduced from the local gray level $I_{\mbox{gray}}$ of the image and the linear relationship between the two quantities obtained beforehand during a calibration step.  Using the following equation, density is measured from the light intensity of the image:

\begin{equation}
    \rho(x,y,t) = \rho_f - \frac{C(x,y,t)}{C_0} (\rho_f - \rho_v),
\end{equation}
with $C(x,y,t) = \alpha I_{\mbox{gray}}$ the fluoresceine concentration at any position and $C_0$ the initial concentration \cite{balasubramanian2018entrainment}. The coefficient $\alpha$ is then used to measured the density field in each experimental run.

\begin{figure}[t]
\centering
\includegraphics[width=13cm]{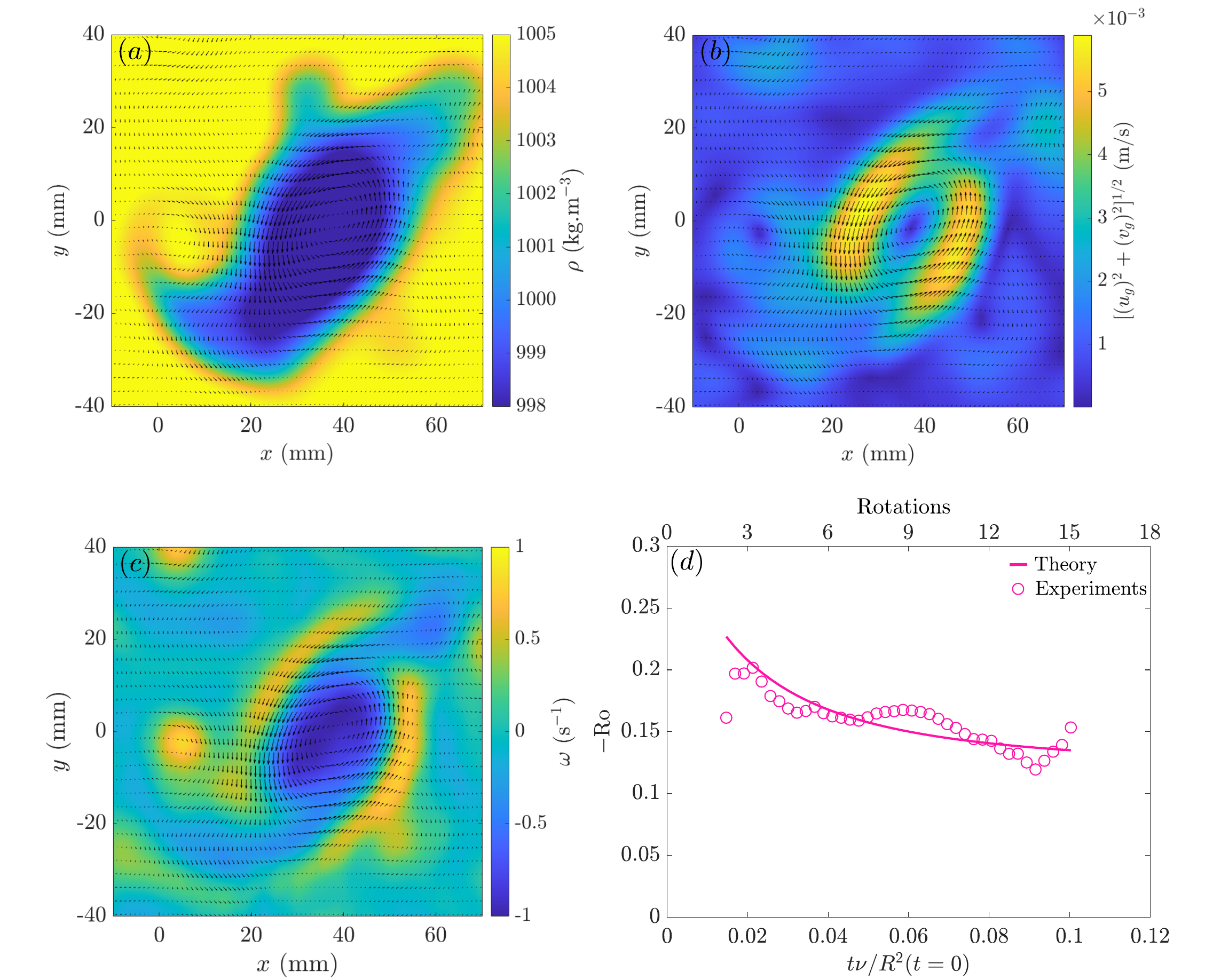}
\caption{Measured (a) horizontal density field $\rho$ and (b) horizontal geostrophic velocity $\vec{u_g}$ magnitude at a fixed vertical position $z_0 = -1$ mm in the vortex for $\textup{Ro} \simeq -0.15$. (c) Corresponding vorticity field $\vec{\omega} \cdot \vec{z} = \vec{\nabla}_h \times \vec{u_g}$. (d) Experimental Rossby number $\textup{Ro} = \omega_c / 2f$ as a function of dimensionless time $t \nu / R^2(t=0)$ where $R^2(t=0)$ is the characteristic radius of the vortex estimated at $t = 0$, compared to the theory of Facchini and Le Bars \cite{facchini2016lifetime} (bottom axis : normalized by the viscous time, top axis : number of table rotations).}
\label{fig:figure5}
\end{figure}

\bigbreak
Once vortices are generated, we wait  10 rotations for the vortex Rossby numbers to decrease (figure~\ref{fig:figure5} (d)). In this regime, measurements of the density and velocity fields are performed as presented in figure~\ref{fig:figure5} (a) and (b). Fresh water is concentrated within the borders of the vortex. A sharp transition between low density (fresh water) and high density (salted water) is visible at the borders of each structures. Within these borders, the velocity field present a stagnation point at the center surrounded by a crown of maximum velocity. In terms of vorticity, the vortex core is an anticyclonic region ($\omega <0$) surrounded by a cyclonic ring ($\omega > 0$) as shown in panel (c) of figure~\ref{fig:figure5}. This is reminiscent of the shielded vortex model presented in section~\ref{sec:shielded}. 

We look at the time evolution of the Rossby number Ro in the vortex presented in figure~\ref{fig:figure5}. The Rossby number is given as a function of the core vorticity $\omega_c$ in our structure and the Coriolis parameter of our experiment, $\textup{Ro} = \omega_c / 2f$. Figure~\ref{fig:figure5} (d) shows $-$Ro as a function of the time normalized by the viscous time scale $R^2/\nu$: $-\textup{Ro}$ decreases with time in the range $0.1 < -\textup{Ro} < 0.2$, meaning our vortex loses intensity with time. This suggests that the Coriolis force is indeed dominant in these dynamics, followed by viscous diffusion and finally non-linear forces.

In the case of a quasigeostrophic and diffusive anticyclonic vortex, Facchini and Le Bars derived an analytical model in the limit of small Rossby and Ekman numbers to explain the time evolution \cite{facchini2016lifetime}. In this case, the Rossby number decreases with time due to viscous diffusion as follows:
\begin{equation}
    \textup{Ro}(\tau) = \frac{\textup{Ro}_0}{\left(1 + 4 \tau \frac{N^2}{f^2}\right)^2},
    \label{eq:eq_Rossby}
\end{equation}
with $\tau = t\nu/R^2(t=0)$ the normalized time and $R(t=0)$ the characteristic radius of the vortex at $t = 0$. Equation~\ref{eq:eq_Rossby} is in excellent agreement with our experimental Ro as shown in figure~\ref{fig:figure5} (d) for $ \textup{Ro}_0 = 0.22$ and $R(t=0) = 0.02$ m.  

\begin{figure}[t]
\centering
\includegraphics[width=14cm]{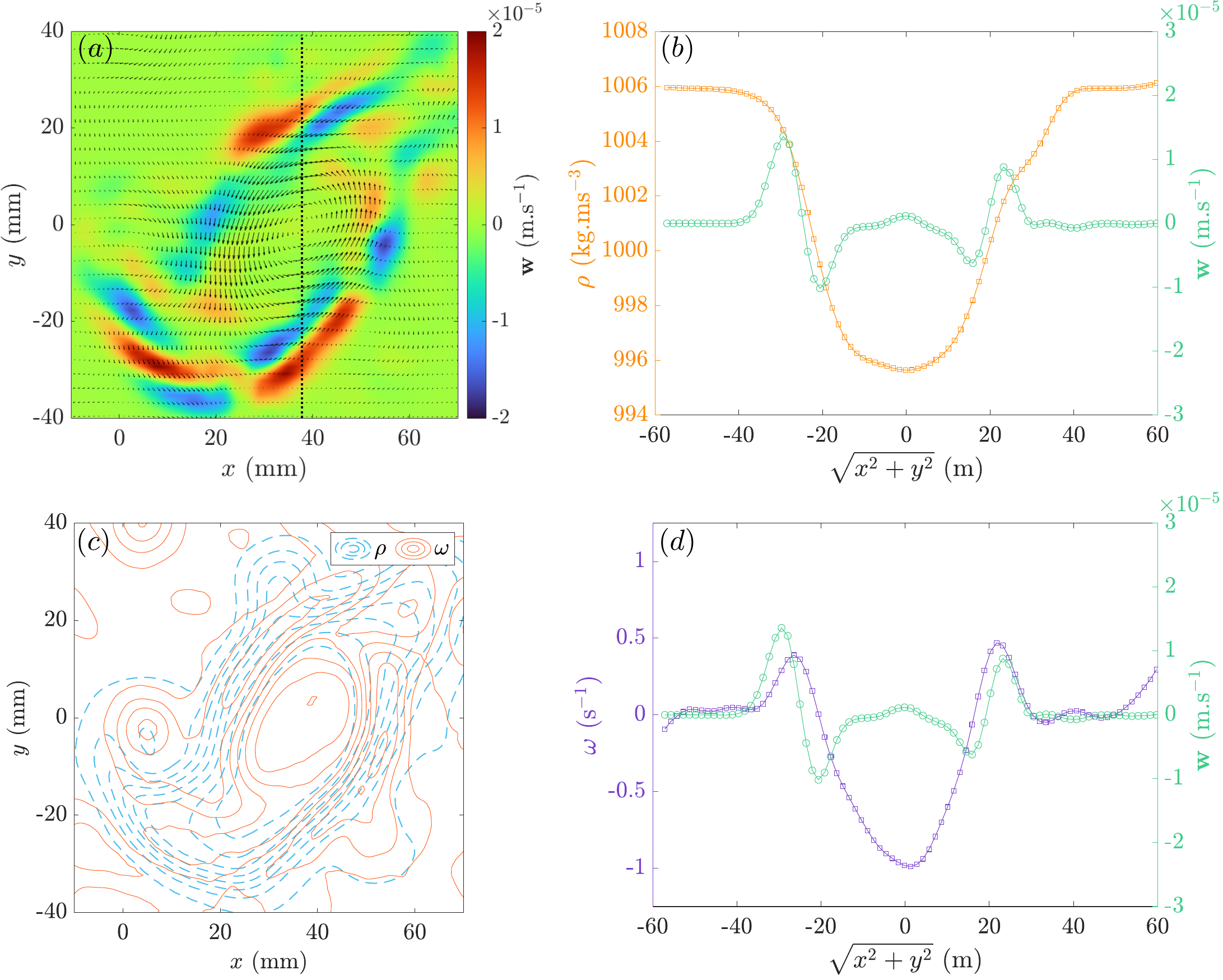}
\caption{At a fixed vertical position $z_0 = -1$ mm below the free surface and $\textup{Ro} \simeq 0.15$, in the vortex : (a) Vertical velocity w estimated with the $\omega-$Equation model.  (b) Profile of vertical velocity w compared with the profile of density $\rho$ extracted from figure~\ref{fig:figure6} (a) along the line $x = 38$ mm passing through the center of the vortex. (c) Isocontours of density $(--)$ and vorticity $(-)$ extracted from figure~\ref{fig:figure6} (a) and (c). (d) Profile of vertical velocity w compared with the profile of vorticity $\omega$ extracted from figure~\ref{fig:figure6} (c) along the line $x = 38$ mm passing through the center of the vortex.}
\label{fig:figure6}
\end{figure}

\subsection{Vertical velocity w measurements}
We now turn to vertical velocity measurements in our set-up. Two methods are used to experimentally reconstruct the vertical component of the velocity field. These will be compared to the $\omega-$Equation diagnosis.

Firstly, as horizontal measurements of $\rho$ and $\vec{u_g}$ are performed at several $z$ positions in the vortex, we can apply equation~\ref{eq:omega_equation}. The resulting vertical velocity is shown for $z_0 = -1$ mm below the free surface in figure~\ref{fig:figure6}. Similarly to the modelled shielded vortex presented in figure~\ref{fig:figure2}, elongated structures of w appear along the edges of the vortex. The maximum vertical velocity is $2 \times 10^{-5}$ m.s$^{-1}$, which is, as expected, quite small compared to the geostrophic velocity ($\simeq 10^{-2}$ m.s$^{-1}$). Figure~\ref{fig:figure6} (b) and (d) show profiles of vertical velocity superimposed with vorticity and density profiles along a line going through the vortex center. As can be observed in panel (b), upward and downward motions appear in the density jump between the fresh water inside the vortex and the salted water outside. This density front is close to the position where vorticity changes sign. Therefore, vertical velocity w appears at the transition between the anticyclonic vorticity core and the cyclonic shield surrounding it (see figure~\ref{fig:figure6} (a) and (d)). Finally, as already discussed, figure~\ref{fig:figure6} (c) shows that vertical velocities arise where density and vorticity isocontours intersect.

\begin{figure}[t]
\centering
\includegraphics[width=13.5cm]{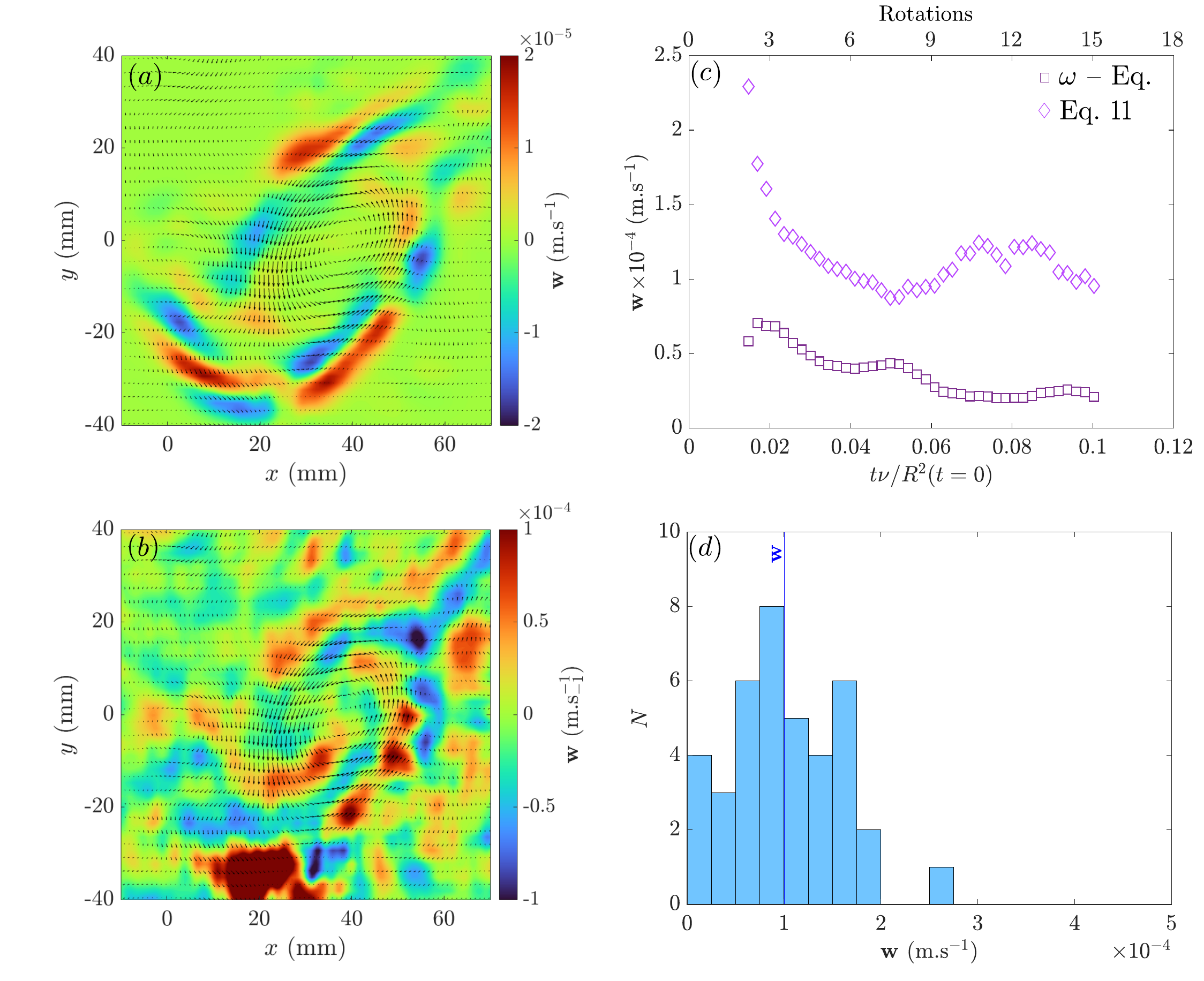}
\caption{At a fixed vertical position $z_0 = -1$ mm below the free surface and $\textup{Ro} \simeq 0.15$, in the vortex : (a) Vertical velocity w estimated with the $\omega-$Equation model. (b) Vertical velocity w estimated from the horizontal divergence (equation~\ref{eq:divergence_w}). (c) Vertical velocity extracted from the horizontal divergence ($\diamond$) and from the $\omega-$Equation prediction ($\square$) as a function of time (bottom axis : normalized by the viscous time $R^2/\nu$, top axis : number of table rotations). (d) Distribution of vertical velocities w estimated from the particle residence time in the laser sheet. The average value of the distribution is indicated in blue $\textup{w} \simeq 10 \times 10^{-5}$ m.s$^{-1}$.}
\label{fig:figure7}
\end{figure}

 Secondly, horizontal divergence of the horizontal velocity field measured through PIV is used to deduce the vertical component, similarly to the ``area rate of change method" used by oceanographers \cite{harlander2012reconstruction,molinari1975calculations}. Applying horizontal divergence to our velocity measurements gives us the vertical gradient of w at each depth $z$. Therefore, using the continuity equation, we can write at first order the vertical velocity w$_i$ at each level $z_i$ chosen every millimeter as

\begin{equation}
    \textup{w}_{i} (\vec{x}) = \textup{w}_{i-1} (\vec{x}) - \int_{z_{i-1}}
^{z_i} \left[ \frac{\partial u}{\partial x} + \frac{\partial v}{\partial y} \right] dz = \textup{w}_{i-1}(\vec{x}) + \left[ \frac{\partial u}{\partial x} + \frac{\partial v}{\partial y} \right]_{z_i} (z_{i-1} - z_i),
\label{eq:divergence_w}
\end{equation}
with the boundary condition $\textup{w}(z = 0) = 0$, i.e. the non penetration condition at the free surface. Our results are presented in figure~\ref{fig:figure7} (a) for a given position below the free surface $z_1 = -1$ mm. We observe a similar pattern to the one obtained with the $\omega-$Equation shown in figure~\ref{fig:figure6} (a) with elongated structures of w along the edges of the vortex. However, in our experiments, the vertical velocity obtained from the divergence method is $10 \times 10^{-5}$ m.s$^{-1}$, which is five times larger than the $\omega-$Equation prediction. 
The time evolution of the $\omega-$Equation prediction and the measurements extracted from the horizontal divergence are presented in figure~\ref{fig:figure7} (b). Both decrease in time and increase with Ro. However, the divergence estimation remains larger than the prediction at all times. If we were to follow the same trend as \cite{viudez2003vertical}, all the estimations of w should converge as Ro decreases because the $\omega-$Equation is better suited for lower Rossby number.

\bigbreak
Finally, a second direct measurement of w is performed by measuring particles residence time in the horizontal laser sheet: vertical velocity forces particles to enter and exit the finite volume of interest. In the region of interest, the laser sheet thickness is on average equal to $e = 0.9 \pm 0.2$ mm. Spatiotemporal images of the particle positions are used to measure the residence time $T$ and, as a consequence, the vertical velocity. Note that this method gives the magnitude but not the sign of the vertical velocity. A distribution of the resulting vertical velocity $\textup{w} = e / T$ is shown in figure~\ref{fig:figure7}  (d) for the same parameters as the divergence estimate in figure ~\ref{fig:figure7}(b). The distribution is centered around an average vertical velocity $\textup{w} \simeq 10 \pm 5 \times 10^{-5}$ m.s$^{-1}$ in agreement with the divergence method, but it is still five times larger than the value obtained from the $\omega-$Equation. Our findings are in agreement with the results of \cite{pietri500skills}. This suggest that the $\omega-$Equation in its current form does not entirely capture the dynamics at play in our experimental system.

\section{Conclusion and perspectives} \label{sec:conclusion}
The main objective of this study was to test the predictive skill of the $\omega$-Equation against laboratory experiments.

In oceanography, the $\omega-$Equation is classically derived under quasi-geostrophic assumptions, and it is of great utility to infer small vertical velocity fields in the ocean from horizontal measurements of submesoscale structures. We have first recalled the main steps required to obtain the $\omega-$Equation from the momentum and mass-conservation equations. Two models of oceanic vortices are then used to examine the predictions of the $\omega-$Equation. As is well known, perfectly axisymmetric vortices do not generate vertical motions. In contrast, as soon as the vorticity contours intersect isopycnal lines, vertical velocities appear. As expected, the magnitude of these vertical velocities is three orders of magnitude smaller than that of the horizontal velocities. This ratio explains why direct measurements at sea are exceedingly difficult.

In a second stage, we set up a laboratory apparatus to generate anticyclonic, floating vortices at the surface of a rotating layer with a Rossby number between -0.3 and -0.1. Using PIV and LIF techniques, we were able to characterize our primarily two-dimensional flow with high accuracy. These measurements of the velocity and density fields were then used to compute the source term of the $\omega-$Equation. The model indeed predicts the generation of a vertical velocity field localized along sharp gradients or fronts surrounding the vortices.

Although the general patterns look like expected ones, the magnitude of the predicted vertical velocity field is smaller by a factor between 3 and 5 to the values we measured independently by two direct methods. Even if the ratio between the predicted and measured values of w decreases with the Rossby number, our results do not confirm the general agreement observed by Viudez and Dritschel on an idealized vortex generated numerically ~\cite{viudez2003vertical}. Note that,  recent measurements of vertical velocities performed in the Ligurian sea current show a similar discrepancy with a more intense in-situ vertical velocity field compared to the $\omega-$Equation predictions  \cite{comby2022measuring,comby2023vitesses}. As already pointed previously, this is also the case described by Pietri et al. \cite{pietri500skills}.

As explained in section~\ref{sec:experiments}, the decrease of the Rossby number in our experiments is in agreement with the dynamical model of Facchini and Le Bars \cite{facchini2016lifetime}. This analytical model states that the combination of the Coriolis force and the viscous diffusion is responsible for the emergence of a vertical recirculation. The viscous diffusion of the azimuthal component of velocity generates a radial current that induces a vertical motion because of flow incompressibility. Finally, this secondary vertical recirculation advects the density anomaly and is responsible for the time evolution of the vortex. Therefore, we can conclude that viscosity may enhance vertical velocities, explaining the difference between our measurements and the $\omega-$Equation. 

Following this hypothesis, we have derived a new version of the $\omega-$Equation including the momentum diffusion through the dynamical viscosity $\nu$ and the scalar diffusion through the diffusivity $\kappa$. The new equation reads :

\begin{equation}
    N^2 \nabla_h^2 \textup{w} + f^2 \frac{\partial^2 \textup{w}}{\partial z^2} = 2 \frac{g}{\rho_0} \vec{\nabla}_h \cdot \left( \vec{\nabla}_h  \vec{u_g} \cdot \vec{\nabla}_h \rho \right) + \frac{g}{\rho_0} (\nu - \kappa) \nabla^2 \nabla_h^2 \rho.
    \label{eq:new_om_eq}
\end{equation} 

\begin{figure}[t]
\centering
\includegraphics[width=13cm]{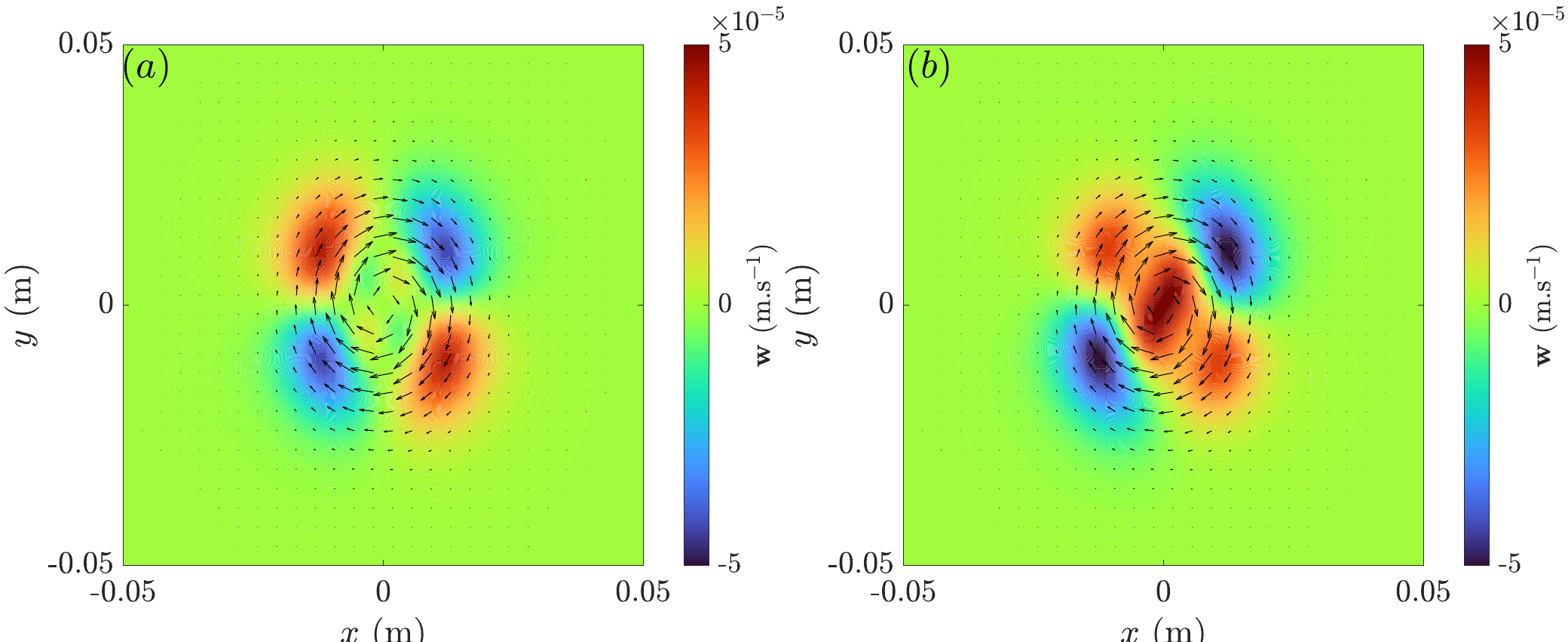}
\caption{For $\textup{Ro} = -0.3$, $r_m = 2 \times 10^{-2}$ m.s$^{-1}$, $f = 2$ s$^{-1}$ and $\rho_0 = 1005$ kg.m$^{-3}$: diagnosis of the vertical velocity w of a Gaussian vortex model with the $\omega-$equation at $\tilde{z} \simeq -0.002$. (a) Estimation using the classical QG $\omega-$Equation formulation (eq. \ref{eq:omega_equation}). (b) Estimation using the new diffusive $\omega-$Equation formulation proposed in equation~\ref{eq:new_om_eq}.}
\label{fig:figure8}
\end{figure}

An additional source term appears involving both diffusion processes. It is interesting to note that momentum diffusion and scalar diffusion compete. In particular, when these processes are equivalent, this new term vanishes and we recover the classical QG $\omega-$Equation: this could be the case when considering turbulent diffusive processes. But in the case of salt diffusion, as it is the case in the ocean and in our experiments, the Schmidt number $Sc = \nu / \kappa$ is 700, so only the viscous term remains. Evaluating orders of magnitude of each source term on the right side on equation~\ref{eq:new_om_eq}, in the conditions of our experiments, shows that the classical and the new viscous source terms may contribute equally to the dynamics of vertical velocities. Unfortunately, the high order of the $\rho$-derivative in the source term makes it extremely difficult to solve the full equation with the fourth order derivatives using experimental data because of the measurement noise inherent to any  experiment. However, figure~\ref{fig:figure8} (b) shows the resulting vertical velocity w with this new formulation tested on the gaussian vortex model presented in section~\ref{sec:omega}. Vertical velocities indeed appear enhanced with this additional diffusive source term. This new version of the $\omega-$Equation opens perspectives for further experimental works for the prediction of vertical velocity field in laboratory generated vortices. 

\section*{Acknowledgments}
This work received financial support from the French government under the France
2030 investment plan, as part of the Initiative d’Excellence d’Aix–Marseille Université – AMIDEX AMX21-RID-035. We thank Stéphanie Barillon, Anne Petrenko and Andrea Doglioli for their collaboration in this project. We thank Uwe Harlander, Louis Gostiaux, Jérôme Noir, Sarah Gille and Stefan Llewellyn-Smith for fruitful discussions. We thank William Le Coz, Eric Bertrand, Adrien Bavière, Léo Knapp and Mathieu Léger for their technical help in constructing the experimental set-up. 

\bibliography{SEALAB}
\bibliographystyle{unsrt}

\end{document}